\documentstyle[epsfig,floats,preprint,prl,aps]{revtex}
\input psfig.sty
\begin{document}
\draft
\title{Measurements of Relative Phase in Binary Mixtures of
  Bose-Einstein Condensates}
\author{D.~S. Hall, M.~R. Matthews, C.~E.  Wieman, and E.~A.
  Cornell\cite{qpdNIST}}
\address{JILA, National
  Institute of Standards and Technology and Department of Physics,\\
  University of Colorado, Boulder, Colorado 80309-0440}
\date{May 21, 1998}
\maketitle
\begin{abstract}
  We have measured the relative phase of two Bose-Einstein condensates
  (BEC) using a time-domain separated-oscillatory-field condensate
  interferometer. A single two-photon coupling pulse prepares the
  double condensate system with a well-defined relative phase; at a
  later time, a second pulse reads out the phase difference
  accumulated between the two condensates. We find that the
  accumulated phase difference reproduces from realization to
  realization of the experiment, even after the individual components
  have spatially separated and their relative center-of-mass motion
  has damped.
\end{abstract}

The relative quantum phase between two Bose-Einstein condensates is
expected to give rise to a variety of interesting behaviors, most
notably those analogous to the Josephson effects in superconductors
and superfluid ${}^3{\mathrm{He}}$~\cite{Backhaus97}. Experiments with
condensates realized in the dilute alkali
gases~\cite{Anderson95,Davis95,Bradley97a} have recently drawn
considerable theoretical attention, with a number of papers addressing
schemes~\cite{Javanainen96b,Imamoglu97a,Ruostekoski97b}
by which to measure the relative phase. Two independent condensates
are expected to possess~\cite{Barnett96} (or develop upon
measurement~\cite{Javanainen96a,Castin97}) a relative phase which is
essentially random in each realization of the experiment. The
experimental observation at MIT of a spatially uniform interference
pattern formed by condensates released from two independent
traps~\cite{Andrews97x} is consistent with this view. In this Letter,
we use an interferometric technique to measure the relative phase (and
its subsequent time-evolution) between two trapped condensates for
which the relative phase is initially \emph{well-defined}. This system
permits us to characterize the effects of couplings to the environment
on the coherence~\cite{coherencenote} between the condensates.

The apparatus and general procedure we use to attain BEC in Rb are
identical to those of the more recent~\cite{Matthews98,Hall98b} of our
previous two-condensate studies~\cite{Matthews98,Hall98b,Myatt97}. We
load roughly $10^9$ atoms in the $\left|F=1, m_F=-1\right>$
($\left|1\right>$) spin state of ${}^{87}$Rb into a time-averaged,
orbiting potential (TOP) magnetic trap~\cite{Petrich95}. We then
magnetically compress and evaporatively cool the gas for 30~s,
producing a condensate of approximately $5 \times 10^5$ atoms with no
noticeable non-condensate fraction ($>75$\% of the gas is in the
condensate). The rotating magnetic field ($\nu_{\mathrm{AF}} =
1800$~Hz) is then ramped to 3.4~G and the quadrupole gradient to
130~G/cm, resulting in a trap with an axial frequency $\nu_z = 59$~Hz.
The fields are chosen to make the hyperfine transition frequency
nearly field-independent~\cite{Hall98a}. We create the second
condensate by applying a short ($\sim400~\mu$s) two-photon pulse that
transfers 50\% of the atoms ($\frac{\pi}{2}$-pulse) from the
$\left|1\right>$ spin state to the $\left|F=2,m_F=1\right>$
($\left|2\right>$) spin state. The coupling drive consists of a
microwave photon at 6833.6640~MHz and a radiofrequency (rf) photon at
1.0134~MHz; the sum of these frequencies is detuned slightly
($\sim100$~Hz) from the expected transition frequency in our
trap~\cite{gpsnote}. After an evolution time $T$ and an optional
second $\frac{\pi}{2}$-pulse, we release the condensates from the
trap, allow them to expand, and image either of the two density
distributions~\cite{Matthews98}. The post-expansion images preserve
the relative positions and gross spatial features of the condensates
as they were in the trap~\cite{Hall98b,Greene_PVT}.

The evolution of the double condensate system, including the coupling
drive, is governed by a pair of coupled Gross-Pitaevskii equations for
condensate amplitudes $\Phi_1$ and $\Phi_2$:
\begin{equation}
  i\hbar\frac{\partial\Phi_1}{\partial t} = 
  \left(T + V_1 + U_1 + U_{12}\right)\Phi_1
  + \frac{\hbar\Omega(t)}{2}e^{-i\omega_{\mathrm{rf}}t}\Phi_2
\label{gpe}
\end{equation}
and
\begin{equation}
  i\hbar\frac{\partial\Phi_2}{\partial t} = \left(T + V_2 +
    V_{\mathrm{hfs}} + U_2 + U_{21} \right)\Phi_2 +
  \frac{\hbar\Omega(t)}{2}e^{i\omega_{\mathrm{rf}}t}\Phi_1
\end{equation}
where $T = -(\hbar^2/2m)\nabla^2$ is the kinetic energy, $m$ is the
mass of the Rb atom, $V_{\mathrm{hfs}}$ is the magnetic
field-dependent hyperfine splitting between the two states in the
absence of interactions, condensate mean-field potentials are $U_i =
4\pi\hbar^2a_in_i/m$ and $U_{ij} = 4\pi\hbar^2a_{ij}n_j/m$,
$n_i=|\Phi_i|^2$ is the condensate density, and the intraspecies and
interspecies scattering lengths~\cite{Matthews98,Hall98b} are $a_i$
and $a_{ij}=a_{ji}$. For the trap parameters given above, the harmonic
magnetic trapping potentials $V_1$ and $V_2$ are displaced from one
another by $0.4~\mu$m along the axis of the trap~\cite{Hall98a}. The
coupling drive is represented here in the rotating wave approximation
and is characterized by the sum of the microwave and rf frequencies,
$\omega_{\mathrm{rf}}$, and by an effective Rabi frequency
$\Omega(t)$, where
\begin{equation}
  \Omega(t)=\cases{2 \pi \cdot 625~{\mathrm{Hz}},& coupling drive
    on;\cr 0, &coupling drive off.}
\end{equation}
Phase-sensitive population transfer between the $\left|1\right>$ and
$\left|2\right>$ states occurs with the drive on, but the two
condensates become completely distinguishable once the drive is
switched off.

\begin{figure}[p]
\begin{center}
\psfig{figure=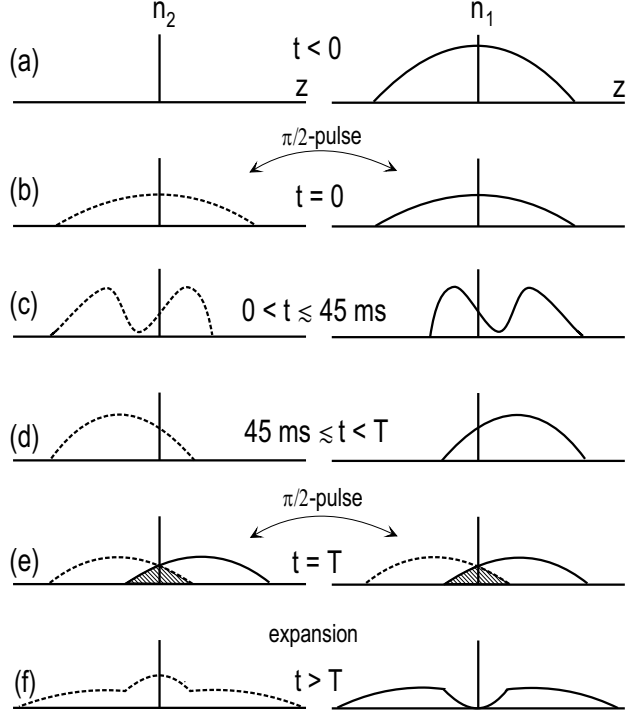,width=0.5\linewidth,clip=}
\end{center}
\caption{A schematic of the condensate interferometer. (a) The
  experiment begins with all of the atoms in condensate
  $\left|1\right>$ at steady-state. (b) After the first
  $\frac{\pi}{2}$-pulse, the condensate has been split into two
  components with a well-defined initial relative phase. (c) The
  components begin to separate in a complicated fashion due to mutual
  repulsion as well as a $0.4~\mu$m vertical offset in the confining
  potentials (see also Fig.~3 of Ref.~\protect\cite{Hall98b}). (d) The
  relative motion between the components eventually damps with the
  clouds mutually offset but with some residual overlap. Relative
  phase continues to accumulate between the condensates until (e) at
  time $T$ a second $\frac{\pi}{2}$-pulse remixes the components; the
  two possible paths by which the condensate can arrive in one of the
  two states in the hatched regions interfere.  (f) The cloud is
  released immediately after the second pulse and allowed to expand
  for imaging. In the case shown, the relative phase between the two
  states at the time of the second pulse was such as to lead to
  destructive interference in the $\left|1\right>$ state and a
  corresponding constructive interference in the $\left|2\right>$
  state.}
\label{expt}
\end{figure}

The first $\frac{\pi}{2}$-pulse [Fig.~\ref{expt}(b)] creates the
$\left|2\right>$ condensate with a repeatable and well-defined
relative phase with respect to the $\left|1\right>$ condensate at
$t=0$. The relative phase between the two condensates subsequently
evolves at a rate proportional to the local difference in chemical
potentials between the two condensates $\omega_{21}(\vec{r},t)$, which
in general is a function of both time and space. Couplings to the
environment~\cite{environmentnote} can induce an additional (and
uncharacterized) precession of the relative phase, leading to an rms
uncertainty in its value
$\Delta\varphi_{\mathrm{diff}}$~\cite{Leggett91,otherdiffnote}.  After
an evolution time $T$, therefore, the condensates have accumulated a
relative phase $\int_0^T\omega_{21}(r,t)\,dt +
\Delta\varphi_{\mathrm{diff}}(T)$. During the same time, the coupling
drive accumulates a phase $\omega_{\mathrm{rf}}T$. A second
$\frac{\pi}{2}$-pulse [Fig.~\ref{expt}(e)] then recombines the
$\left|1\right>$ and $\left|2\right>$ condensates, comparing the
relative phase accumulated by the condensates to the phase accumulated
by the coupling drive. The resulting phase-dependent beat note is
manifested in a difference in the condensate density between the two
states. Immediately after the second pulse the density in the
$\left|2\right>$ state ($n_{2f}$) is
\widetext
\begin{equation}
  n_{2f}(\vec{r}) =
  \frac{1}{2}n_1(\vec{r})+\frac{1}{2}n_2(\vec{r})
+\sqrt{n_1(\vec{r})n_2(\vec{r})}
  \cos\left[\left(\int_0^T\omega_{21}(\vec{r},t)\,dt\right) - \omega_{\mathrm{rf}}T
    + \Delta\varphi_{\mathrm{diff}}(T)\right].
\label{phase}
\end{equation}
\narrowtext
In this equation, $n_i$ denote the densities prior to the
application of the second $\frac{\pi}{2}$-pulse. The interference
term in Eq.~\ref{phase} shows that measurement of $n_{2f}(\vec{r})$ in
the overlap region is sensitive to the relative phase. Each
realization of the experiment (with a freshly-prepared condensate)
yields a measurement of the relative phase for a particular $T$; by
varying $T$, we can measure the evolution of the relative phase.

\begin{figure}[p]
\begin{center}
\psfig{figure=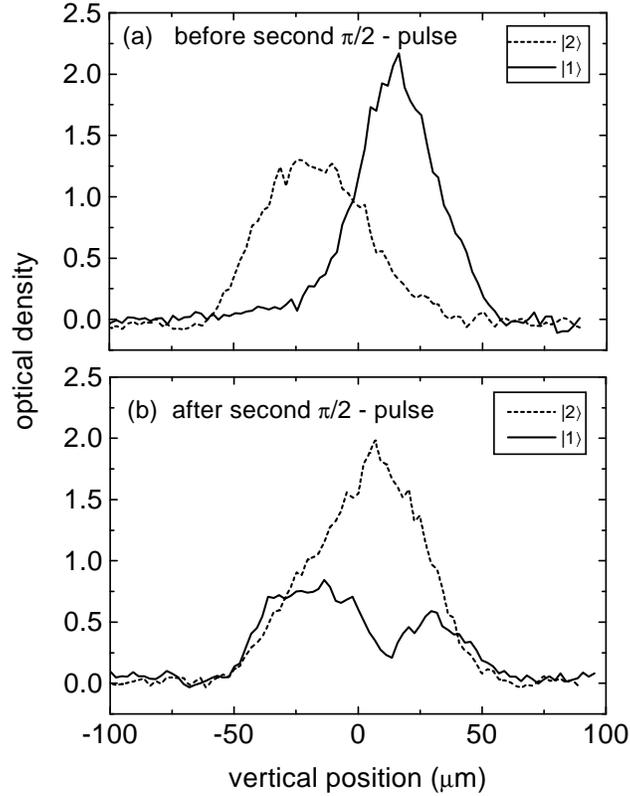,width=0.5\linewidth,clip=}
\end{center}
\caption{(a) The post-expansion density profiles of the condensates
  in the steady-state attained after a single $\frac{\pi}{2}$-pulse.
  These density profiles vary little from shot-to-shot (and
  day-to-day). (b) The density profiles after the second
  $\frac{\pi}{2}$-pulse. The density in the overlap region depends on
  the relative phase between the two condensates at the time of the
  pulse; in the case shown, we observe constructive interference in
  the $\left|2\right>$ state and destructive interference in
  $\left|1\right>$. The patterns in (b) are much less stable than
  those in (a), possibly as a result of unresolved higher-order condensate
  excitations, issues associated with the expansion, or technical
  instabilities of the apparatus.}
\label{reprise}
\end{figure}

At short times $T$, for which the overlap between the condensates
remains high, varying the moment at which the second
$\frac{\pi}{2}$-pulse is applied causes an oscillation of the
total resulting number of atoms in the $\left|2\right>$ state. The oscillation
occurs at the detuning frequency $\delta = \omega_{21}-\omega_{\mathrm{rf}}$
and is completely analogous to that observed in
separated-oscillatory-field measurements in thermal atomic
beams~\cite{Ramsey56} or in cold (but noncondensed) atoms in a
magnetic trap~\cite{Hallunpub}. The fringe contrast, initially 100\%,
decreases as the condensates separate. After $\sim45$~ms the relative
center-of-mass motion damps and comes to equilibrium, leaving the
components with a well-defined overlap region at their boundary, as
shown in Figs.~\ref{expt}(d) and~\ref{reprise}(a); see also Fig.~5(b)
of Ref.~\cite{Hall98b}.  Application of a second $\frac{\pi}{2}$-pulse
at $T \gtrsim45$~ms results in a density profile in which the
interference occurs only in the overlap region; see
Figs.~\ref{expt}(f) and~\ref{reprise}(b).

\begin{figure}[p]
\begin{center}
  \psfig{figure=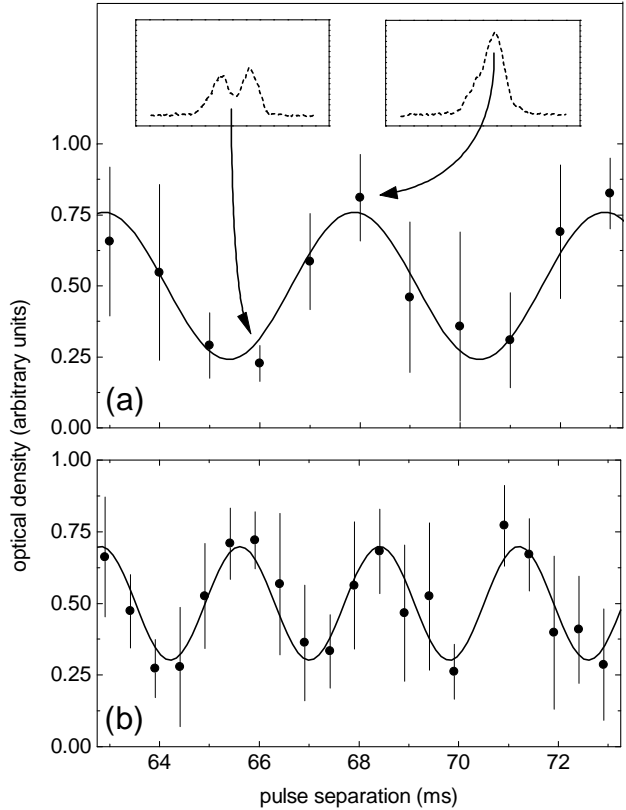,width=0.5\linewidth,clip=}
\end{center}
\caption{The value of the condensate density in the
  $\left|2\right>$ state is extracted at the center of the overlap
  region (inset) and plotted (a) as a function of $T$. Each point
  represents the average of 6~separate realizations and the thin bars
  denote the rms scatter in the measured interference for an
  individual realization.  The thick lines are sinusoidal fits to the
  data, from which we extract the angular frequency
  $\omega_{21}-\omega_{\mathrm{rf}}$. In (b), the frequency of the
  coupling drive $\omega_{\mathrm{rf}}$ has been increased by
  $2\pi\cdot150$~Hz, leading to the expected reduction in fringe
  spacing. }
\label{latephi}
\end{figure}

We look at the density of atoms in the $\left|2\right>$ state at the
center of the overlap region~\cite{overlapnote} in order to
examine the intriguing issue of the reproducibility of the relative
phase accumulated by the condensates during the complicated approach
to equilibrium. If the phase diffusion term in Eq.~\ref{phase} is so
large that the uncertainty is greater than $\pi$, then repeated
measurements for the same values of $T$ will yield an incoherent
(\emph{i.e.}, random) ensemble of interference patterns. In the
opposite extreme, (\emph{i.e.}, very little phase diffusion), repeated
measurements will give essentially the same interference pattern at
$T$ in each experimental run. We plot the optical density in the
center of the overlap region as a function of $T$ in
Fig.~\ref{latephi}, and observe an oscillation at the detuning
frequency with a visibility of approximately 50\%, corresponding to an
rms phase diffusion $\Delta\varphi_{\mathrm{diff}}(T) \lesssim 
\frac{\pi}{3}$.  At longer times the maximum contrast observed in a single
realization of the experiment decreases slightly, possibly due to the
increasing presence of thermal atoms as the condensates decay.

The stable interference patterns show that the condensates retain a
clear memory of their initial relative phase despite the complicated
rearrangement dynamics of the two states following the first
$\frac{\pi}{2}$-pulse. This is rather surprising, since the
center-of-mass motion of the double condensate system is strongly (and
completely) damped, and, in general, decoherence times in entangled
states tend to be much shorter than damping
times~\cite{Caldeira85,Walls85,Brune96}. The intuition one develops in
understanding few-particle quantum mechanics may not apply to
experiments involving condensates. The phase
between the two condensates seems to possess a robustness which preserves
coherence in the face of the ``phase-diffusing'' couplings to the
environment.

We have read out the relative phase of two Bose-Einstein condensates
using a time-domain version of the method of separated oscillatory
fields. We observe the persistence of phase memory in
this ``condensate interferometer,'' despite the presence of damping
and the complicated rearrangement of the two condensate components. We
have established that the time scale for phase diffusion in this
system can be longer than 100~ms. The double-condensate methods we have
developed will be applicable to other experiments which explore phase
diffusion as a function of condensate parameters including
temperature, number of atoms~\cite{Wright96,Javanainen97,Lewensteinq},
and collision rates~\cite{Wong96}. Collapses and revivals of the
``memory'' of the relative phase are
predicted~\cite{Castin97,Wright96} at time scales which may be
experimentally accessible should environmentally-induced diffusion
effects remain small. Our methods will also allow us to examine other
phase-related phenomena, such as phase-locking and analogues of the
superconducting Josephson junctions~\cite{Javanainen86q}.

We gratefully acknowledge useful conversations with A.~J. Leggett, as
well as with the other members of the JILA BEC collaboration.  This
work is supported by the ONR, NSF, and NIST.


\begin{thebibliography}{10}

\bibitem[*]{qpdNIST}
Quantum Physics Division, National Institute of Standards and Technology.

\bibitem{Backhaus97}
S. Backhaus {\it et~al.}, Science {\bf 278},  1435  (1997).

\bibitem{Anderson95}
M.~H. Anderson {\it et~al.}, Science {\bf 269},  198  (1995).

\bibitem{Davis95}
K.~B. Davis {\it et~al.}, Phys. Rev. Lett. {\bf 75},  3969  (1995).

\bibitem{Bradley97a}
C.~C. Bradley, C.~A. Sackett, and R.~G. Hulet, Phys. Rev. Lett. {\bf 78},  985
  (1997).

\bibitem{Javanainen96b}
J. Javanainen, Phys. Rev. A {\bf 54},  R4629  (1996).

\bibitem{Imamoglu97a}
A. Imamo{\=g}lu and T.~A.~B. Kennedy, Phys. Rev. A {\bf 55},  R849  (1997).

\bibitem{Ruostekoski97b}
J. Ruostekoski and D.~F. Walls, Phys. Rev. A {\bf 56},  2996  (1997).

\bibitem{Barnett96}
S.~M. Barnett, K. Burnett, and J.~A. Vaccaro, J. Res. Natl. Inst. Stand.
  Technol. {\bf 101},  593  (1996).

\bibitem{Javanainen96a}
J. Javanainen and S.~M. Yoo, Phys. Rev. Lett. {\bf 76},  161  (1996).

\bibitem{Castin97}
Y. Castin and J. Dalibard, Phys. Rev. A {\bf 55},  4330  (1997).

\bibitem{Andrews97x}
M.~R. Andrews {\it et~al.}, Science {\bf 275},  637  (1997). Due to technical
  noise, the intrinsically random nature of the interference pattern was not
  conclusively established.

\bibitem{coherencenote}
  
  We define ``coherence'' as the predictability of the relative
  quantum phase. For interesting discussions of quantum coherence, see
  Refs.~\protect\cite{Leggett95,Zurek91}.

\bibitem{Leggett95}
A.~J. Leggett,  in {\em {B}ose-{E}instein Condensation}, edited by A. Griffin,
  D.~W. Snoke, and S. Stringari (Cambridge University Press, Cambridge, 1995).

\bibitem{Zurek91}
W.~H. Zurek, Phys. Today {\bf 44} (10),  36  (1991).

\bibitem{Matthews98}
M.~R. Matthews {\it et~al.}, e-print cond-mat/9803310. Submitted to
PRL as ``Dynamical Response of a Bose-Einstein Condensate to a Discontinuous
Change in Internal State.''

\bibitem{Hall98b} 
 D.~S. Hall M.~R. Matthews, J.~R. Ensher, C.~E.  Wieman, and E.~A.
 Cornell, e-print cond-mat/9804138. Submitted to PRL as ``The
 Dynamics of Component Separation in a Binary Mixture of
 Bose-Einstein Condensates.''

\bibitem{Myatt97}
C.~J. Myatt {\it et~al.}, Phys. Rev. Lett. {\bf 78},  586  (1997).

\bibitem{Petrich95}
W. Petrich, M.~H. Anderson, J.~R. Ensher, and E.~A. Cornell, Phys. Rev. Lett.
  {\bf 74},  3352  (1995).

\bibitem{Hall98a}
D.~S. Hall {\it et~al.}, Proc. SPIE {\bf3270} (in press).

\bibitem{gpsnote}
The microwave and rf frequencies are produced by synthesizers locked to
  global-positioning system (GPS) signals. The manufacturer of the receiver
  claims a root Allan variance better than $10^{-10}$.

\bibitem{Greene_PVT}
C.~H. Greene, (private communication).

\bibitem{environmentnote} We separate the ``environment'' into two
  categories: an ``intimate'' environment, which includes interactions
  with thermal atoms as well as with internal modes within the
  condensate itself; and an ``external'' environment which includes
  such experimental factors as uncontrolled fluctuations in the
  magnetic fields. The former are intrinsic to the physics of the
  problem, whereas the latter can (in principle) be suppressed. In
  practice, we experience difficulty in keeping the external
  environment from intruding on our measurements; only in the modes of
  quietest operation are the oscillations of
  Fig.~\protect\ref{latephi} observable. When coherence is observed,
  perturbations due to \emph{both} the intimate and the external
  environments must be small, whereas when coherence is washed out,
  \emph{either} may be responsible. For these reasons, we have not
  attempted in this paper to quantify the loss of coherence at longer
  times.

\bibitem{Leggett91}
A.~J. Leggett and F. Sols, Found. Phys. {\bf 21},  353  (1991).

\bibitem{otherdiffnote} Quantum fluctuations introduce an additional
  uncertainty that grows linearly in time in a process akin to the
  spreading of a Gaussian wavepacket in space; see
  Refs.~\protect\cite{Castin97,Wright96,Javanainen97,Lewensteinq}. We
  estimate the time scale for this process to be much longer than the
  duration of an individual measurement.


\bibitem{Wright96}
E.~M. Wright, D.~F. Walls, and J.~C. Garrison, Phys. Rev. Lett. {\bf 77},  2158
   (1996).

\bibitem{Javanainen97}
J. Javanainen and M. Wilkens, Phys. Rev. Lett. {\bf 78},  4675  (1997).

\bibitem{Lewensteinq}
M. Lewenstein and L. You, Phys. Rev. Lett. {\bf
    77}, 3489 (1996); A. Imamo{\=g}lu, M. Lewenstein, and L. You,
  Phys. Rev. Lett. {\bf 78}, 2511 (1997).

\bibitem{Ramsey56}
N.~F. Ramsey, {\em Molecular Beams} (Clarendon Press, Oxford, 1956).

\bibitem{Hallunpub}
D.~S. Hall {\it et~al.}, (unpublished).

\bibitem{overlapnote}
We take the average of a $\sim14~\mu$m wide (post-expansion) vertical swath
  down the middle of the condensate density profile and extract the amplitude
  of the pixel at the center of the condensate image (\emph{i.e.}, at the
  center of the overlap region).

\bibitem{Caldeira85}
A.~O. Caldeira and A.~J. Leggett, Phys. Rev. A {\bf 31},  1059  (1985).

\bibitem{Walls85}
D.~F. Walls and G.~J. Milburn, Phys. Rev. A {\bf 31},  2403  (1985).

\bibitem{Brune96}
M. Brune {\it et~al.}, Phys. Rev. Lett. {\bf 77},  4887  (1996).

\bibitem{Wong96}
T. Wong, M.~J. Collett, and D.~F. Walls, Phys. Rev. A {\bf 54},  R3718  (1996).

\bibitem{Javanainen86q}
J. Javanainen, Phys. Rev. Lett. {\bf 57},  3164  (1986);
I. Zapata, F. Sols, and A.~J. Leggett, Phys. Rev. A {\bf 57},  R28  (1998);
A. Smerzi, S. Fantoni, S. Giovanazzi, and S.~R. Shenoy, Phys. Rev. Lett. {\bf
  79},  4950  (1997);
J. Williams and M. Holland, (private communication).

\end{thebibliography}

\end{document}